\documentclass[sigconf]{acmart}
\AtBeginDocument{%
  }

\usepackage{booktabs}   
\usepackage{subcaption} 
\usepackage{multirow}
\usepackage{tcolorbox}
\usepackage{listings}
\usepackage{xcolor}
\usepackage[T1]{fontenc}
\usepackage{graphicx}
\usepackage{makecell}

\begin{document}

\title{Beyond a Single Queue: Multi-Level-Multi-Queue as an Effective Design for SSSP problems on GPUs}

\newcommand\codeemph[1]{{\ttfamily #1}}

\newcommand{\circled}[1]{%
\tikz[baseline=(char.base)]{
\node[shape=circle,draw,inner sep=1pt] (char) {#1};}}

\fancyhead{}  
\renewcommand\footnotetextcopyrightpermission[1]{} 

\newcommand{\lowspeedup}[0]{1.87}
\newcommand{\highspeedup}[0]{17.13}
\newcommand{\com}[1]{\textcolor{red}{#1}}

\newcommand{\graybox}[1]{%
\begin{tcolorbox}[
  colback=gray!8,
  colframe=black,
  boxrule=0.4pt,
  arc=2pt,
  left=2pt,
  right=2pt,
  top=2pt,
  bottom=2pt
]
#1
\end{tcolorbox}
}

\author{%
Zhengding Hu\textsuperscript{1},
Jingwei Sun\textsuperscript{1},
Le Jiang\textsuperscript{1},
Yuhao Wang\textsuperscript{1},
Junqing Lin\textsuperscript{1},
Yi Zong\textsuperscript{2},
GuangZhong Sun\textsuperscript{1} \\
\textsuperscript{1}\textit{University of Science and Technology of China} \quad
\textsuperscript{2}\textit{Tsinghua University} \\
}


\begin{abstract}
As one of the most fundamental problems in graph processing, the Single-Source Shortest Path (SSSP) problem plays a critical role in numerous application scenarios. However, existing GPU-based solutions remain inefficient, as they typically rely on a single, fixed queue design that incurs severe synchronization overhead, high memory latency, and poor adaptivity to diverse inputs. To address these inefficiencies, we propose \textbf{Multi-Level-Multi-Queue (MLMQ)}, a novel data structure that distributes multiple queues across the GPU’s multi-level parallelism and memory hierarchy. To realize MLMQ, we introduce a cache-like collaboration mechanism for efficient inter-queue coordination, and develop a modular queue design based on unified Read and Write primitives. Within this framework, we expand the optimization space by designing a set of GPU-friendly queues, composing them across multiple levels, and further providing an input-adaptive MLMQ configuration scheme. Our MLMQ design achieves average speedups of 1.87x to 17.13x over state-of-the-art implementations. Our code is open-sourced at \url{https://github.com/Leo9660/MLMQ.git}.
\end{abstract}

\maketitle

\section{Introduction}

Single-Source Shortest Path (SSSP) is a fundamental problem in computer science, with applications in a wide range of fields, including transportation \cite{delling2009transportation}, bioinformatics \cite{barabasi2004networkbioinfo}, circuit simulation \cite{fishburn2003tiloscircuit}, social networks \cite{kempe2003socialnetworks}, recommendation systems \cite{ying2018recommendation}, and modern data-centric workloads such as information retrieval \cite{malkov2018hnsw} and knowledge graphs \cite{hogan2021knowledgegraph}.

\begin{figure}[t]
    \centering
    \includegraphics[width=\linewidth]{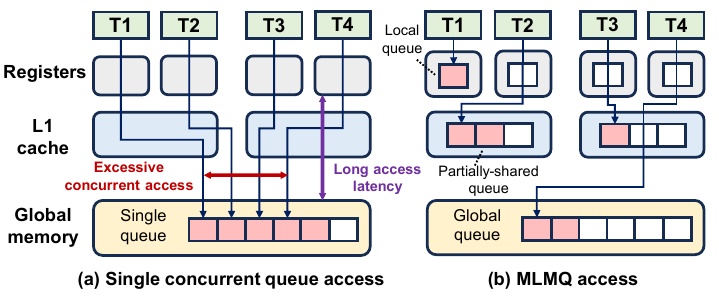}
    \caption{Differences between single concurrent queue and MLMQ. $T_i$ refers to the $i$-th GPU thread.}
    \label{pic:introview}
\end{figure}


A typical SSSP algorithm is associated with a \emph{queue} (also termed as a \emph{scheduler} in some studies) for task management and scheduling. Each iteration dequeues a vertex and performs relaxing operations on it. Relaxing refers to updating the shortest known distances to a vertex's neighbor when a shorter path is discovered. The behavior of the queue determines the total workload and runtime efficiency.


Over the decades, designing efficient queues for SSSP has remained a central challenge. Research ranges from theoretical designs \cite{Dijkstra, fredman1987fibonacciheap, meyer2003delta, haeupler2024universal, duan2025barrier} to practical CPU implementations and optimizations \cite{shun2013ligra, dhulipala2017julienne, dong2021cpuefficient, lenharth2015prioritynotgood, yesil2019understanding}. With the growing prevalence of GPUs in modern systems, GPU-based queue designs have attracted substantial attention. Many studies have explored concurrent queues on GPUs, including relaxed priority queues \cite{alistarh2015spraylist, moscovici2017gpu, he2012design, crosetto2019cupq, chen2021bgpq}, concurrent FIFO queues \cite{burtscher2012relatedquantitative, busato2015hbf, surve2017relatedparallel}, and multi-bucket queues \cite{davidson2014work, wang2021fast}. However, these studies face a common design limitation, namely the reliance on a \textit{Single, Global, Fixed-type concurrent queue}. This limitation gives rise to the following performance challenges:


\textbf{\emph{Excessive Synchronization under Massive Parallelism}.} Compared with traditional multi-core CPUs, a GPU has significantly higher parallelism. As such, the well-established methods to maintain concurrent queues on multi-core CPUs \cite{lenharth2015prioritynotgood, rihani2015multiqueues, yesil2019understanding, postnikova2022multi, nguyen2013obim1, nguyen2011obim2} become less efficient, since simultaneous accesses of thousands of threads and warps can lead to high synchronization overhead.
    
\textbf{\emph{Bandwidth Bottleneck of Global Memory Access}.} 
GPUs demand high data efficiency to match the massive parallelism. However, concurrent queues accessible to all threads are typically maintained in global memory, turning limited global bandwidth into a bottleneck and leaving higher-level memory such as the L1 cache and registers underutilized.

\textbf{\emph{Limited Adaptability of Single-Type Queues}.} Parallel execution improves speedup at the cost of work efficiency: the optimal sequential order \cite{Dijkstra, haeupler2024universal} is relaxed, leading to redundant enqueues and relaxations in exchange for greater parallelism. Striking the right balance between parallelism and work efficiency is thus of critical importance. Achieving this balance is particularly challenging because no single queue type is universally effective: different queue types can exhibit varying performance on different graphs \cite{dong2021spaa}. This makes input-adaptive adjustment of queue policies a necessity for robust performance.



To this end, we propose a novel data structure design, \emph{Multi-Level-Multi-Queue} (shortened as MLMQ). As the name suggests, MLMQ maintains multiple queues on multiple levels of parallelism and memory hierarchy. Figure \ref{pic:introview} illustrates the distinction between MLMQ and the single, global queue. By introducing local and partially shared queues, MLMQ reduces the frequency of concurrent accesses to a single queue and significantly mitigates synchronization overhead. Furthermore, distributing queue elements across hierarchical memory lowers access latency and alleviates the pressure on global bandwidth. 
Finally, each level of the queue can be treated as a modular design, enabling flexible combinations of different queue types for better input adaptability.



Nevertheless, implementing the concept of MLMQ is non-trivial, due to three primary challenges: \textbf{First}, coordinating queues across registers, shared memory, and global memory is difficult, since GPUs lack efficient inter-warp and inter-block communication. CPU-side techniques like work stealing \cite{blumofe1999stealing, postnikova2022multi} do not directly apply. \textbf{Second}, maintaining local queues can result in frequent priority violations, as each GPU warp tends to access data from higher-level queues rather than strictly following the global highest priority. Experimental results show that using a basic FIFO policy for higher-level queues can cause performance to degrade by as much as 31\%, due to an additional workload of 45\%. \textbf{Third}, MLMQ expands the design space by combining different queue types across multiple levels, making it challenging to select configurations that balance parallelism, work efficiency, and data locality for diverse graphs.



To address the above challenges, we introduce an inter-level data exchange and queue collaboration mechanism with low-cost cache-style data transfers. The mechanism is unified through a set of MLMQ Read and Write primitives, enabling modular queue abstractions that support flexible configurations. Building on the primitives, we implement a variety of GPU queues, from refined FIFO policies to lightweight partially ordered variants, each exploring different trade-offs between access cost and work efficiency. We then systematically combine these queues on different levels and develop an input-adaptive strategy that automatically selects efficient configurations for different graph types.


In summary, the contributions of the paper are as follows.

\begin{itemize}
\item We propose MLMQ, an novel data structure design to solve SSSP problems on GPUs, which coordinates multiple queues across multi-level parallelism and memory hierarchies. 
\item We develop a framework based on MLMQ that provides modular queue design primitives and interfaces. Within this framework, we implement a variety of GPU queues across different levels, with an input-adaptive strategy to select effective queue configurations for diverse graph inputs.
\item Evaluation on real-world graphs demonstrates that our MLMQ-based framework achieves average speedups of \lowspeedup $\times$ to \highspeedup $\times$ over state-of-the-art baselines, providing an effective and comprehensive solution for SSSP problems on GPUs.
\end{itemize}

\section{Background and Motivation}
\subsection{Single-Source Shortest Path}

\begin{figure}[t]
    \centering
    \includegraphics[width=0.95\linewidth]{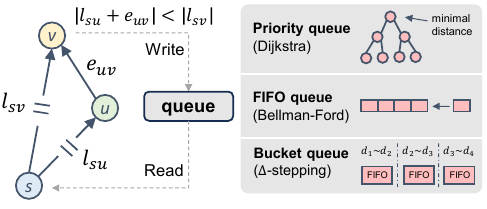}
    \caption{Relaxing and enqueue/dequeue operations in SSSP. Different queues correspond to different algorithms.}
    \label{pic:relax example}
    \vspace{-0.5em}
\end{figure}

Given a weighted graph $G(V, E, W)$, where $V$ is the set of vertices, $E = \{e_{uv} | u, v \in V \}$ is the set of edges, 
the Single-Source Shortest Path (SSSP) problem aims to find the set of shortest paths starting from a source vertex $s$: $L(s) = \{l_{su} | u\in V\}, s\in V$, where $l_{su}$ indicates the shortest path from $s$ to $u$. $|\cdot|$ to denote the length of an edge. 
\par
As shown in Figure \ref{pic:relax example}, a typical SSSP algorithm maintains a queue. In each iteration, the algorithm reads a vertex $u$ from the queue and performs relaxing operations on it. The \emph{relaxing} operation traverses all the neighboring vertices in $N(u) = \{v |\ e_{uv}\in E\}$ and examines whether $\{l_{su}, e_{uv}\}$ forms a shorter path to $v$. The vertex $v$ will be written back to the queue when $l_{sv}$ is updated. The algorithm starts with vertex $s$ in the queue and iterates until the queue is empty. 


\textbf{Endpoints of Parallelism and Work Efficiency}. The most classic SSSP algorithms are Dijkstra's algorithm \cite{Dijkstra} and Bellman-Ford algorithm \cite{BellmanFord}. Dijkstra's algorithm maintains a priority queue, where the vertex with the highest priority is dequeued and relaxed in each iteration. Although work efficiency of this algorithm is optimal \cite{haeupler2024universal}, the strict prioritization limits the available parallelism. Conversely, the Bellman-Ford algorithm maintains a simple FIFO queue. All vertices in the queue can be relaxed simultaneously without considering priority. This algorithm offers high parallelism but at the expense of significantly redundant operations. The two algorithms are the two endpoints of high work efficiency and high parallelism, respectively.

\textbf{Intermediate Queue Designs for Balanced Trade-offs}. On modern parallel processors, an ideal balancing point between the above two extremes leads to better utilization of the hardware resources. 
For example, Multi-queue \cite{rihani2015multiqueues} maintains several priority queues and associates each one with different parallel units. Such a design can reduce resource contention and improve data locality. 
$\Delta$-stepping algorithm \cite{meyer2003delta} maintains a multi-bucket queue. Each bucket stores the vertices with priorities ranging in a given interval $\Delta$. Vertices are relaxed in parallel in each bucket. As such, parallelism in vertices with similar priorities can be exploited, while work efficiency is ensured by processing buckets in ascending order of priority ranges. Scalable queue design in distributed scenarios \cite{wang2022scaling, gan2021tianhegraph, yesil2019understanding, lin2018shentu, nguyen2013lightweight, nguyen2011synthesizing} has also been extensively studied, mostly based on multi-core CPUs.

\subsection{Single Queue Performance Bottlenecks}
\label{sec:motivation}

\begin{figure}[t]
    \centering
    \includegraphics[width=\linewidth]{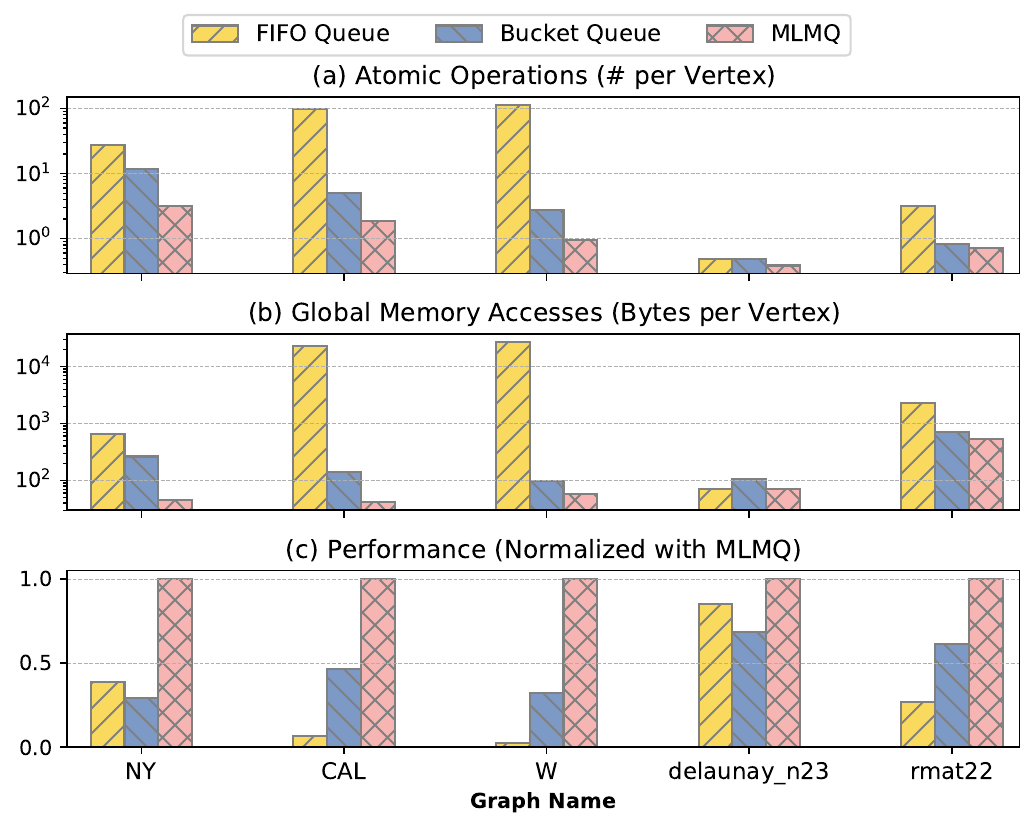}
    \caption{Performance profiling with different queues on NVIDIA 3080 Ti. Results are normalized by the number of vertices, which corresponds to the optimal work, i.e., the number of relaxations in Dijkstra’s algorithm.}
    \label{pic:moti-type}
\end{figure}

Existing SSSP algorithms generally employ a single-queue design \cite{Dijkstra, BellmanFord, rihani2015multiqueues, meyer2003delta}. On GPUs, the single queue typically resides in global memory and is concurrently accessed by all the threads \cite{davidson2014work, busato2015hbf, wang2021fast}. Such an access pattern is poorly aligned with the massive parallelism and multi-level memory hierarchy of GPUs. We conduct performance profiling of two GPU-friendly queues: the FIFO queue from Bellman-Ford algorithm and the bucket queue from $\Delta$-stepping algorithm. As shown in Figure~\ref{pic:moti-type}, we identify three critical performance bottlenecks of the single queue: 



\graybox{
\noindent \textbf{\emph{Bottleneck 1:  Highly Concurrent Queue Access Brings \\ Excessive Parallel Contention.}}
}

SSSP exhibits an inherently high degree of concurrent queue access on GPUs. In each iteration, a large number of threads simultaneously dequeue active vertices, traverse their outgoing edges, and conditionally enqueue neighboring vertices whose distances are updated. As the wavefront expands, these enqueue and dequeue operations become highly concurrent, causing frequent accesses to shared queue elements and metadata.

Under a single-queue design, queue maintenance is serialized through a small set of shared variables, including queue sizes,  head and tail pointers, as well as queue elements. To preserve correctness, these shared states must be protected by atomic operations. As a result, the amount of synchronization grows rapidly with parallelism, often far exceeding the amount of effective computation.

Figure~\ref{pic:moti-type}(a) shows that the total number of atomic operations substantially surpasses the optimal work, reaching up to 11.8$\times$ for FIFO queues and 99.1$\times$ for bucket queues. 
The contention on these shared structures forces GPU threads to serialize, significantly limiting achievable parallelism and wasting available compute resources.




\graybox{\noindent \textbf{\emph{Bottleneck 2: Frequent Global Memory Access Brings\\ High-Latency Queue Operations.}}}

In existing GPU implementations, a single queue is typically placed in global memory to enable visibility to all the threads. As a result, enqueue and dequeue operations repeatedly incur high-latency global memory accesses. Unlike arithmetic operations or on-chip memory accesses, these accesses cannot be efficiently hidden when they dominate the execution.

As shown in Figure~\ref{pic:moti-type}(b), the number of global memory accesses can exceed the effective work by up to four orders of magnitude. This excessive memory traffic arises not only from accessing queue elements themselves, but also from repeatedly loading and updating queue-related metadata. The resulting long memory access latency and bandwidth pressure significantly degrade overall efficiency.

\graybox{\noindent \textbf{\emph{Bottleneck 3: A Single Queue Type Only Provides Limited Adaptability to Diverse Input Graphs.}}}

A single queue design cannot effectively accommodate the diversity of real-world input graphs, which exhibit substantial variations in sparsity patterns, degree distributions, and structural properties. As shown in Figure~\ref{pic:moti-type}(c), different queue types demonstrate clear performance preferences: the FIFO queue performs better on mesh-like graphs, while the bucket queue is more effective on power-law graphs. Moreover, even within the same graph category, such as road networks, the optimal queue choice varies with graph scale. These results indicate that no single queue structure consistently delivers robust performance across inputs, highlighting the need for queue design that can adjusts to varying graph characteristics.


In comparison, our solution (MLMQ) significantly reduces both the synchronization overhead and the global memory access burden. Its input-adaptive configuration ensures consistently strong performance under varying input graphs.

\subsection{Challenges for Multiple Queues on GPUs}
\label{sec:moti-multiqueue}

Given the substantial performance bottlenecks introduced by the single-queue design (\S \ref{sec:motivation}), a natural question arises: can multiple queues be employed on multiple memory levels, to increase parallelism and reduce access overhead? 
On CPU systems, multiple-queue structures have been extensively studied \cite{rihani2015multiqueues, postnikova2022multi, lenharth2015prioritynotgood, yesil2019understanding, nguyen2013obim1, nguyen2011obim2}, but are mostly confined to single-level platforms with shared memory or point-to-point (P2P) communication.
In contrast, the GPU's hierarchical architecture and programming model complicate the multiple-queue design, summarized as follows:



\graybox{\noindent \textbf{\emph{Challenge 1: GPU Programming Model and Memory Hierarchy Complicate Multi-Queue Coordination.}}}

Coordinating multiple queues on GPUs is complicated by the visibility, accessibility, and capacity constraints imposed by the memory hierarchy. For performance, queue operations are ideally served from fast on-chip storage. However, queues placed in registers or shared memory are inherently confined to a thread block: other blocks cannot directly access these queue elements, which makes global coordination difficult. In contrast, placing queues in global memory leads to high-latency operations.

The GPU memory hierarchy further complicates multi-queue coordination due to its capacity–latency tradeoff. For example, on an NVIDIA A100 GPU, shared memory provides at most 64--164$\sim$KB per SM, whereas traversal wavefronts, defined as the set of vertices to be relax in each iteration step in large graphs commonly exceed $10^5$ vertices. Such wavefronts therefore cannot be kept in fast on-chip memory and must be managed in global memory. As a result, an efficient design must carefully manage large wavefronts in the global memory to maintain scalability, while effectively exploiting limited on-chip memory only for lower access costs.

\begin{figure}[t]
    \centering
    \includegraphics[width=\linewidth]{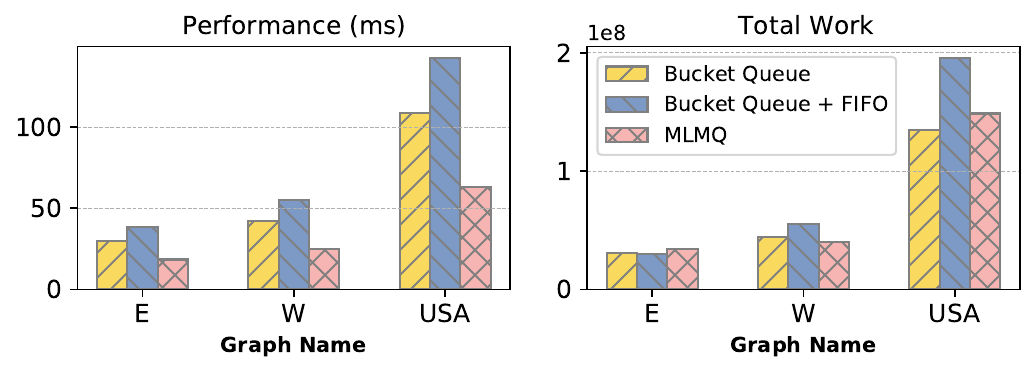}
    \caption{Performance and total work comparison when using local queues on a bucket queue.}
    \label{pic:priorityinversion}
\end{figure}

\graybox{\noindent \textbf{\emph{Challenge 2: Priority Inversion Brings Redundant \\Wavefronts, Leading to Lower Work Efficiency.}}}

When maintaining local queues, the issue of priority inversion is exacerbated by the GPU's massive degree of parallelism. Figure \ref{pic:priorityinversion} illustrates that naively assigning a private queue to each thread results in up to 31\% performance degradation, because each thread prefers processing local tasks instead of high-priority ones, with high frequent priority inversions and up to 45\% additional work.

\begin{figure}[t]
    \centering
    \includegraphics[width=0.7\linewidth]{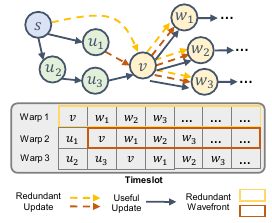}
    \caption{An example of redundant wavefronts.}
    \label{pic:redundantwave}
\end{figure}

Upon further analysis, we observe that the private queue leads to the emergence of \emph{redundant wavefronts}, a phenomenon where a single redundant update triggers a cascade of further redundant updates. As illustrated in Figure~\ref{pic:redundantwave}, assume the neighbors ($v$, $u_1$, $u_2$) of a source vertex $s$ are assigned to three different warps, and the edge weights satisfy $|e_{sv}| > |e_{su_1}| + |e_{u_1v}| > |e_{su_2}| + |e_{u_2u_3}| + |e_{u_3v}|$. Warp 1 first performs a redundant update on $v$, then proceeds to relax its subsequent vertices ($w_*$) based on $|l_{sv}| = |e_{sv}|$. Warp 2 later relaxes $u_1$ and also performs a redundant update on $v$, continuing to relax downstream vertices using $|l_{sv}| = |e_{su_1}| + |e_{u_1v}|$. Only Warp 3 performs the truly useful update on $v$. However, since Warp 1 and 2 reach each subsequent vertex earlier, they will initiate a series of redundant updates, forming multiple wavefronts that propagate sequentially across the graph. Such an inefficient pattern substantially degrades work efficiency, particularly on large-diameter graphs.

\graybox{\noindent \textbf{\emph{Challenge 3: Multi-Queue Enlarges the Optimization Space.}}}

Introducing multiple queues significantly enlarges the design space, particularly along the dimension of memory placement. Queues residing at different levels of the GPU memory hierarchy obey fundamentally different design constraints, including capacity limits, access latency, and synchronization scope. While each queue type can be efficient in isolation, combining queues across memory levels into a unified design that simultaneously balances parallelism, data locality, and work efficiency is non-trivial.

This challenge is further exacerbated by the input-sensitive nature of SSSP. Introducing multiple queues significantly amplifies configuration sensitivity, as different graphs favor different combinations of queue types, placements, and parameters. As a result, the enlarged design space becomes difficult to tune statically, since configurations that are effective for some queues or inputs are often suboptimal when considered globally.

\section{MLMQ: Design and Method}

\begin{figure}[t]
    \centering
    \includegraphics[width=\linewidth]{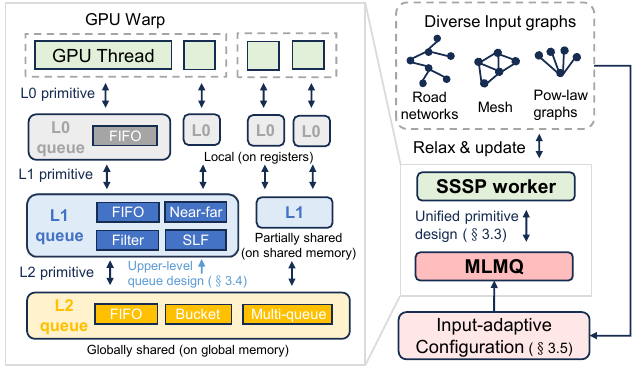}
    \caption{An overview of the MLMQ-based framework.}
    \label{pic:overview}
\end{figure}

In this paper, we propose Multi-Level-Multi-Queue (MLMQ), a novel and efficient queue structure tailored for SSSP problems on GPUs. The overview design of MLMQ is shown in Figure \ref{pic:overview}. By integrating \textbf{multiple queues} with the GPU’s \textbf{multi-level} hierarchy, and carefully orchestrating the collaboration and behavioral patterns across different queues, we effectively address both the performance bottlenecks in single-queue designs (\S\ref{sec:motivation}) and the challenges associated with multiple-queue management (\S \ref{sec:moti-multiqueue}). 

\subsection{Multi-level Queue Structure}

MLMQ adopts a three-level structure. Each level corresponds to one level of GPU parallelism and memory hierarchy. The descriptions of three levels of queues are as follows:

\textbf{L0 queue} is the private queue. Each running thread of the SSSP worker has its own L0 queue. The data is stored and managed with registers, enabling minimum-cost access. The capacity of the L0 queue is constrained by the total number of available GPU registers.

\textbf{L1 queue} is the partially-shared queue. Each warp (32 threads on NVIDIA GPUs) shares one L1 queue. The data is stored in shared memory, involving relatively low access costs due to the utilization of on-chip memory. The capacity of the L1 queue is constrained by the amount of shared memory available to each warp.

\textbf{L2 queue} is the globally-shared queue. All threads share the L2 queue. The queue data is stored and shared in global memory, which means higher access costs due to the long latency of global memory requests and the demands for synchronization. While slower, the L2 queue offers larger capacity compared to L0 and L1 queues, with an upper bound equal to the available GPU device memory.

In summary, L0, L1, L2 queues are stored in registers, shared memory, and global memory, and used by a thread, a thread warp, and the whole grid, respectively. By distributing queue data across all three levels, MLMQ makes effective use of the upper-tier, low-latency memory hierarchy. Threads and warps can efficiently fetch data from private or partially shared queues, alleviating pressure on global memory bandwidth and reducing synchronization costs.

\subsection{Cache-like Queue Collaboration}

\definecolor{lightgray}{gray}{0.97}
\definecolor{codegray}{gray}{0.4}
\definecolor{codeblue}{rgb}{0.25,0.35,0.75}

\definecolor{codegreen}{rgb}{0,0.6,0}
\definecolor{codegray}{rgb}{0.5,0.5,0.5}
\definecolor{codepurple}{rgb}{0.58,0,0.82}
\definecolor{backcolour}{rgb}{0.95,0.95,0.92}
\definecolor{textblue}{rgb}{.2,.2,.7}
\definecolor{textred}{rgb}{0.54,0,0}
\definecolor{textgreen}{rgb}{0,0.43,0}
\definecolor{codered}{rgb}{201,72,12}

\lstdefinestyle{cudadsl}{
  language=C++,
  basicstyle=\ttfamily\footnotesize,   
  backgroundcolor=\color{lightgray},
  frame=single,
  rulecolor=\color{gray!30},
  frameround=tttt,
  numbers=left,
  numberstyle=\tiny\color{codegray},
  numbersep=5pt,
  breaklines=true,
  keepspaces=true,
  showstringspaces=false,
  columns=fixed,   
  commentstyle=\color{textred}\itshape,
  stringstyle=\color{textgreen},
  keywordstyle=\bfseries\color{textblue},
  literate={Element}{{{\color{textblue}\bfseries Element}}}7%
           {Read}{{{\color{codegreen}\bfseries Read}}}4%
           {Write}{{{\color{codegreen}\bfseries Write}}}5%
}

\begin{figure}[t]
\begin{minipage}{\linewidth}
\begin{lstlisting}[style=cudadsl, caption={MLMQ Read and Write primitives. The data type of elements in the queue is defined as \codeemph{Element}. The boolean flag valid is used for the thread predicate.}, label={list:rwprim}]
// Queue pointers stored in registers
L0_queue *q0; L1_queue *q1; L2_queue *q2;
...

Status Read(Element &input, bool &valid) {
  if (q0->Read(input, valid) == READ_EMPTY)
    if (q1->Read(input, valid) == READ_EMPTY)
      if (q2->Read(input, valid) == READ_EMPTY)
        return READ_EMPTY;
  return SUCCESS;
}

Status Write(Element output, bool valid) {
  // Buffers for write-back elements in L0 and L1 queue
  __shared__ Element *b1, *b2; int n1, n2;
  if (q0->Write(output, valid, b1, n1) == WRITE_BACK)
    if (q1->Write(b1, n1, b2, n2) == WRITE_BACK)
      q2->Write(b2, n2);
  return SUCCESS;
}
\end{lstlisting}
\end{minipage}
\end{figure}


MLMQ enables effective collaboration across multiple queues via data exchange between adjacent levels. The SSSP worker performs reads and writes only on the top-level queue (the L0 queue). When the L0 queue is empty or full, it triggers a read or a write operation of the L1 queue. Similarly, when the L1 queue becomes empty or full, it triggers a read or a write operation of the L2 queue.

Such collaboration design follows two intuitions. First, it provides an efficient approach for data exchange between private and partially shared queues, namely, through read and write operations to lower-level queues. Compared to some CPU-side techniques such as work stealing \cite{blumofe1999stealing, postnikova2022multi}, the approach is more suitable for GPUs with lower synchronization overhead and contention risks. Second, the data exchange pattern between adjacent queues is similar to that of data movement in multi-level GPU cache systems. Such close alignment with hardware behavior brings optimization opportunities without introducing additional overhead beyond the inherent cost of memory accesses.

\subsection{Modular Queue with Unified Primitives}

Built upon the cache-like collaboration mechanism, MLMQ adopts a modular queue design with unified primitives for versatility and generality. As illustrated in Listing \ref{list:rwprim}, MLMQ exposes two unified primitives to the SSSP worker: Read and Write, representing queue data access operations along with corresponding queue management. The internal behavior and management of MLMQ are determined by its modular L0, L1, and L2 queues. Each level implements its Read and Write primitives within the designated memory layer to perform data access and queue management. The primitives are executed in a warp-level collaborative manner, transferring a batch of elements each time to prevent warp divergence and maximize memory coalescing. 

Through unified primitives, MLMQ enables modular queue design and development, allowing different implementations to be easily integrated at any level while simplifying implementation and maintenance. This modularity also provides the flexibility to configure queues across levels, making it easier to adapt the queue to different input graphs.




\subsection{Upper-level Queue Design}

The design of each level of queues has a significant impact on performance. For the L2 queue, we provide efficient concurrent implementations (\S \ref{sec:ds cons}) based on existing algorithms, including FIFO queue \cite{BellmanFord}, bucket queue \cite{meyer2003delta}, and priority queues \cite{Dijkstra, rihani2015multiqueues}. For the L0 and L1 queues, given their limited capacity and critical influence on work efficiency, we introduce specialized designs for queue behaviors, specifically the semantics of the Read/Write primitives.

\textbf{L0 queue.}
The capacity of the L0 queue is strictly limited by the number of GPU registers. Moreover, the instruction flow of threads in L0 Read/Write primitives should be more consistent to avoid warp divergence. Therefore, the L0 queue is simply implemented as a FIFO vector, aiming to buffer elements and reduce the number of accesses to lower-level queues. When warp-level Read primitive is invoked, data is read from thread-private queues and evenly scattered across threads using warp primitives. When the L0 queue is full, all its elements are transferred to the L1 queue. 

\begin{figure}[t]
    \centering
    \includegraphics[width=\linewidth]{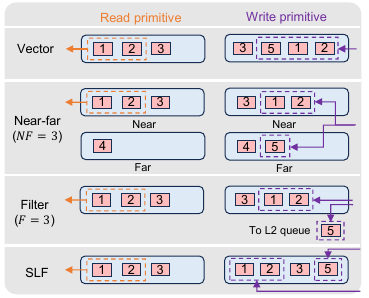}
    \caption{The illustration of different L1 queues. Suppose the L1 Read primitive reads 2 elements, and the L1 Write primitive is invoked with elements with distance 5, 1, 2.}
    \label{pic:l1 queue}
\end{figure}

\textbf{L1 queue.} GPU L1 shared memory has larger capacity than registers and is well-suited for warp-level parallel access. We design four types of L1 queues, as illustrated in Figure \ref{pic:l1 queue}. These variants provide distinct trade-offs between access latency, management overhead, and the extent to which priority is preserved (referred to as \emph{priority guarantee}).

\emph{L1 Vector} is designed as a FIFO vector. The queue is completely priority-unaware, but incurs the lowest management overhead, as Read and Write primitives are operated through sequential insertion and removal of the elements. To mitigate the issue of priority inversion, we introduce a periodic flushing strategy: after every $wb$ invocations of the Write primitive, all elements in the FIFO queue are flushed back once to the L2 queue. This strategy ensures that the SSSP worker does not repeatedly operate on local queue data, effectively preventing the emergence of redundant wavefronts.

\emph{L1 Near-far queue} is based on the intuition that elements with priorities close to the optimal are more likely to represent useful work. Therefore, we prefer to relax these elements earlier. The queue consists of two vectors with a threshold $NF$. Elements with the shortest path lengths less than $NF$ are written to the near vector, while those with larger lengths are written to the far vector. When Read primitive is invoked, elements in the near vector are read first. When the near vector is empty, the Read routine sets $NF = min + \Delta$, where $min$ is the minimum shortest path length of elements in the far vector. Then, elements in the far vector are reorganized based on the new $NF$. 

\emph{L1 Filter queue} follows a similar design with the L1 Near-far queue. The difference is that the L1 Filter queue maintains only a single vector along with a filtering threshold $F$. When the Write primitive is invoked, only elements with distances $\le F$ are reserved in the vector, while the rest are written back to the L2 queue. Compared to the Near-far queue, this design introduces more frequent accesses to the L2 queue. However, it ensures the timely write-back of lower-priority vertices, thus offering higher inter-warp parallelism for graphs with less degree of priority inversion.

\emph{L1 Shortest-Length-First queue} is designed for stronger priority guarantees. It maintains a double-ended queue to improve work efficiency. The Read primitive preferentially retrieves elements from the head, while the Write primitive performs a priority comparison—placing elements longer than the head element at the head, and all others at the tail. When the queue size exceeds its capacity, the overflowed elements are written back to the L2 queue. Despite higher access costs due to dynamic comparisons, the queue can effectively mitigate the issue of local priority inversion.

In summary, our MLMQ framework supports multiple queue instances across different levels. The overview of the queues and their design trade-offs is presented in Table~\ref{tab:instance}. At the L1 and L2 levels, MLMQ supports a range of queues with varying degrees of priority guarantees and access efficiency. By combining these queues, the framework is able to construct advanced queue structures tailored with distinct performance characteristics, resulting in performance advantages across diverse graph inputs.

\begin{table}[t]
\centering
\scalebox{0.9}{
\begin{tabular}{ccc
    >{\centering\arraybackslash}p{1.4cm}
    >{\centering\arraybackslash}p{2.1cm}}
    \toprule
    Level & Queue name & Abbr. & \makecell[c]{Priority\\Guarantee} & \makecell[c]{Access\\Efficiency}\\
    \midrule
    \multirow{1}{*}{L0} & Vector & L0V & $\Downarrow$ & $\Uparrow$ \\
    \midrule
    \multirow{3}{*}{L1} & Vector & L1V & $\Downarrow$ & $\Uparrow$ \\
    & Near-far queue & L1NF & $\Rightarrow$ & $\Rightarrow$ \\
    & Filter queue & L1FQ & $\Rightarrow$ & $\Rightarrow$ \\
    & SLF queue & L1SLF & $\Uparrow$ & $\Downarrow$ \\
    \midrule
    \multirow{4}{*}{L2} & Vector & L2V & $\Downarrow$ & $\Uparrow$ \\
    & Bucket queue & L2DQ & $\Rightarrow$ & $\Rightarrow$ \\
    & Priority queue & L2PQ & $\Uparrow$ & $\Downarrow$  $\Downarrow$ \\
    & Multi-queue & L2MPQ & $\Uparrow$ & $\Downarrow$  $\Downarrow$ \\
    \bottomrule
\end{tabular}
}

\caption{Instances of queue at each level}
\label{tab:instance}
\vspace{-1em}
\end{table}

\subsection{Input Adaptive Configuration}
\label{sec:adaptive}

Given the substantial variation in structure and size of real-world graphs, adaptively selecting the optimal MLMQ configuration is crucial for performance consistency. Our configuration strategy is informed by the experimental results in \S\ref{sec:test case}, which reveal the following key insights:

\begin{itemize}
    \item Graph structure has a strong impact on performance: power-law graphs are highly sensitive to priority inversion, whereas meshes show much lower sensitivity.  
    \item Graph scale affects the trade-off between parallelism and work efficiency: smaller graphs typically benefit from parallel-friendly queues, while larger graphs require more work-efficient ones to avoid too much redundant work.  
    \item Mixed-priority-guarantee configurations, with stronger priority at one level and weaker at the other, often provide the best balance between overhead and efficiency.  
\end{itemize}


\begin{table}[t]
\begin{tabular}{ll}
\toprule
\textbf{Feature} & \textbf{Description} \\
\midrule
$\text{m}$              & Number of vertices \\
$\text{nnz}$     & Number of edges \\
$\text{avg\_nnz}$ & Average degree per vertex \\
$\text{max\_nnz}$ & Maximum vertex degree \\
$\text{dev\_nnz}$ & Standard deviation of vertex degree \\
$\text{avg\_weight}$ & Average edge weight \\
$\text{dev\_weight}$ & Standard deviation of edge weight \\
$\text{max\_weight}$ & Maximum edge weight \\
\bottomrule
\end{tabular}
\centering
\caption{Graph features for adaptive configuration.}
\label{tab:features}
\end{table}

Beyond these heuristics, we further implement a learning-based selector to automate adaptive configuration. Specifically, we train a RandomForest regressor~\cite{breiman2001randomforest} to predict the relative performance of candidate queue configurations. During input preprocessing, MLMQ extracts a lightweight set of graph features that are easily available at load time, including vertex and edge counts (capturing graph size), as well as average degree, maximum degree, and degree variance, etc (capturing structural properties). The selected features are summarized in Table~\ref{tab:features}. The model then predicts both the most suitable queue type and the corresponding parameters (e.g., selecting $NF$ for the L1 Near-Far queue or $F$ for the L1 Filter queue). Based on the prediction results, our solution ranks candidate configurations and chooses the one with the highest expected performance, enabling a low-overhead, input-adaptive configuration.

\section{Implementation}
\subsection{MLMQ-based Kernel Implementation}

Our GPU kernel consists of two decoupled components: an \emph{MLMQ manager} and an \emph{SSSP worker}. This separation allows queue management and termination detection to proceed independently of relaxation work. We employ the persistent thread technique, where warps repeatedly execute until all relaxations are completed, avoiding costly kernel relaunches. The number of blocks matches the number of streaming processors to ensure high occupancy.

A key issue is workload imbalance caused by skewed vertex degrees. Assigning one vertex per thread leads to stragglers when degrees vary widely. 
We use two strategies to handle workload imbalance: vertices above a threshold degree are processed cooperatively within a warp to mitigate long tails, while smaller-degree vertices are distributed across threads for balanced load. This hybrid approach achieves both high utilization and fairness.

Our MLMQ-based GPU kernel consists of two decoupled components: an MLMQ manager and an SSSP worker. The pseudocode of the MLMQ-based SSSP algorithm is shown in Listing \ref{list:sssp}. We employ the persistent thread technique wherein a consistent group of threads executes the SSSP code repeatedly until all relaxing operations are completed. Two GPU kernels are launched: \textbf{Work kernel} repeatedly reads vertices from MLMQ (Line 27), performs relaxing operations (Line 11-18), and writes new vertices into MLMQ (Line 19). We refer to the warps in this kernel as working warps. Each working warp is associated with a status (Line 4), indicating whether its execution should terminate. \textbf{Manage kernel} continuously checks the size of MLMQ (Line 34). When MLMQ is empty, termination signals are sent to all the working warps (Line 36-37). We refer to the warp executing this kernel as the manage warp.

\lstdefinestyle{cudacode}{
  language=C++,
  basicstyle=\ttfamily\footnotesize,
  backgroundcolor=\color{lightgray},
  frame=single,
  frameround=tttt,
  rulecolor=\color{gray!30},
  framesep=2mm,
  numbers=left,
  numberstyle=\tiny\color{codegray},
  numbersep=5pt,
  breaklines=true,
  keepspaces=true,
  showstringspaces=false,
  columns=fixed,
  escapeinside={||}, 
  commentstyle=\color{textred}\itshape,
  stringstyle=\color{textgreen},
  keywordstyle=\bfseries\color{textblue},
  morekeywords={__global__,__device__,__host__,__shared__,__forceinline__},
  literate={Element}{{{\color{textblue}\bfseries Element}}}7%
           {Read}{{{\color{codegreen}\bfseries Read}}}4%
           {Write}{{{\color{codegreen}\bfseries Write}}}5%
}

\begin{figure}[t]
  \begin{minipage}{\linewidth}
    \begin{lstlisting}[
      style=cudadsl,
      caption={Parallel SSSP implementation based on MLMQ},
      label={list:sssp}
    ]
#define NUM_WARP (NUM_BLOCK * BLOCK_SIZE / 32)
MLMQ *q; // Stored in registers
struct Element {int id; int dist}; // Queue element
bool work_flag[NUM_WARP]; // Status of each warp
...
__device__ void SSSP_relax(Element u, bool valid) {
  // Eliminate duplicate vertices
  if (valid && u.dist > dist[u.id]) valid = false;
  // Relax all the edges collaboratively
  // edge_list contains all neighbors of valid elements
  for ((uid, vid) |\textbf{in}| edge_list) |\textbf{in thread parallel}| {
    Element new_v;  bool update = false;
    if (edge_weight(uid, vid) + dist[uid] < dist[vid]) {
      new_v.id = vid;
      new_v.dist = edge_weight(uid, vid) + dist[uid];
      update = true;
      dist[vid] = new_v.dist;
    }
    q->Write(new_v, update);
  }
}

__global__ void work_kernel(...) {
  int wid = (blockIdx.x * BLOCK_SIZE + threadIdx.x) / 32;
  Element u;  bool valid;
  while (work_flag[wid] == true) {
    if (q->Read(u, valid) == SUCCESS)
      SSSP_relax(u, valid);
  }
}

__global__ void manage_kernel(...) {
  // Check if MLMQ is empty
  while (!q->is_empty()) {}
  // Set flags of all working warps to false
  for (int w = 0; w < NUM_WARP; w++)
    work_flag[w] = false; 
}

void sssp_on_mlmq() {
  cudaStream_t s1, s2;
  work_kernel<<<NUM_BLOCK, BLOCK_SIZE, s1>>>(...);
  manage_kernel<<<1, 32, s2>>>(...);
}
    \end{lstlisting}
  \end{minipage}
\end{figure}

In each iteration, the working warp reads a batch of vertices $u$ from MLMQ, traverses all their neighbor vertices $v$ and performs relaxing operations. Assigning one vertex $u$ to a single thread can lead to workload imbalance due to uneven vertex degrees. 
We thus adopt two parallel relaxing strategies, drawing on methods from \cite{davidson2014work} and \cite{wang2021fast}. 
For a vertex with a degree exceeding a threshold $TH\_V$, which is set to 16 for Ampere GPUs, all threads within a warp cooperatively process its neighbors to avoid the long tail effect. For vertices with degrees below this threshold, their neighbors are evenly distributed among threads to achieve load balancing.


\begin{figure}[t]
  \begin{minipage}{\linewidth}
    \begin{lstlisting}[
      style=cudadsl,
      caption={Block-based lock-free implementation of concurrent L2 FIFO queue.},
      label={list:l2_vec}
    ]
struct L2_vector {
  int writePtr = 0;
  int readPtr = 0;
  Element block_data[BLOCK_NUM][BLOCK_SIZE];
  int block_size[BLOCK_NUM];
  // Transfer elements from address v to the vector
  Status Write(Element *v, int write_num) {
    int newPtr;
    if (threadIdx.x % 32 == 0)
      newPtr = atmoicAdd(&writePtr, 1) % BLOCK_NUM;
    newPtr = __shfl_sync(0xffffffff, newPtr, 0);
    while (block_size[newPtr] > 0) {}
    // collaboratively copy data
    data_copy(block_data[newPtr], v, write_num);
    block_size[newPtr] = write_num;
    return SUCCESS;
  }
  // Transfer elements from the vector to address v
  Status Read(Element *v, int &read_num)
  {
    __shared__ int newPtr = -1;
    if (newPtr < 0 && threadIdx.x % 32 == 0)
      newPtr = atmoicAdd(&readPtr, 1) % BLOCK_NUM;
    newPtr = __shfl_sync(0xffffffff, newPtr, 0);
    if (block_size[newPtr] > 0) {
      read_num = block_size[newPtr];
      // collaboratively copy data
      data_copy(v, block_data[newPtr], read_num);
      block_size[newPtr] = 0;
      newPtr = -1;
      return SUCCESS;
    }
    else
      return READ_EMPTY;
  }
};
    \end{lstlisting}
  \end{minipage}
\end{figure}

One challenge is judging whether the MLMQ is empty. The Manage kernel only inquire about the size of the L2 queue but remains unaware of the local queue size of the working warp. Thus, we adopt a delayed updating strategy. MLMQ maintains two shared values \codeemph{global\_reserve} and \codeemph{global\_done}. When L2 Write routine is invoked, the value of \codeemph{global\_reserve} is atomically increased by the number of written elements. When L2 Read routine is invoked, the number of read elements is accumulated to a local register \codeemph{local\_done}. Only when L2 Read routine returns \codeemph{READ\_EMPTY}, the value of \codeemph{local\_done} is atomically accumulated into \codeemph{global\_done}. Before the SSSP kernel starts, the source vertex is first written through to the L2 queue. The strategy delays the update of the global queue size until the local queues of the current warp are empty. Therefore, the manage kernel can determine the MLMQ is empty when \codeemph{global\_reserve} equals to \codeemph{global\_done}.


\subsection{Concurrent L2 Queue Implementation}
\label{sec:ds cons}
While L0 and L1 queues can be built with simple warp-level instructions, L2 queues must support highly concurrent access across warps. In this section, we explore several concurrent queue designs relevant to SSSP on GPUs, and present their implementations and optimization strategies when integrated into MLMQ.

\textbf{L2 FIFO queue.} A common approach to implementing a globally shared concurrent FIFO queue is using a mutex lock to manage exclusive read/write. However, this approach incurs considerable synchronization overhead and severely limits parallelism. Here, we implement an efficient FIFO queue based on data blocks instead of single elements. Read and Write primitives are performed at the granularity of a data block to enable coarse-grained memory access. Two global pointers (\codeemph{writePtr} and \codeemph{readPtr}) and one local pointer (\codeemph{readPtr}) are employed for concurrent access, with an array \codeemph{block\_size} to indicate the valid element number inside each block. When Write primitive is invoked, the warp atomically increases \codeemph{writePtr} and transfers elements into the block it points to. When Read primitive is invoked, the warp atomically increases \codeemph{readPtr} and stores the pointer in \codeemph{newPtr} (reside in shared memory). Then the warp repeatedly checks the size of the block that \codeemph{newPtr}, until the target blocks have valid elements to read. Values of atomic operations are propagated from thread 0 to others with shuffle instructions (\codeemph{\_\_shufl\_sync}). The algorithm follows a lock-free design with only one \codeemph{atomicAdd} operation involved at each Read and Write primitive, which ensures the access performance under high concurrency.



\textbf{L2 Bucket queue.} We implement the parallel-friendly queue proposed in $\Delta-$stepping algorithm \cite{meyer2003delta} based on L2 FIFO queues. L2 bucket queue consists of several concurrent L2 FIFO queues, each corresponding to a bucket in the algorithm. The $i-th$ bucket contains elements with priority values in the range of $[base + \Delta * (i-1), base + \Delta * i)$. When Read primitive is invoked, the warp first checks the bucket with the highest priority, and subsequently checks the next bucket if the previous one is empty. The queue has $bmax$ buckets in total and the first $bnum$ buckets are checked. $bnum$ is adjusted to balance work efficiency and parallelism. The value of $base$ is increased by $\Delta$ when the first bucket is empty after the Read primitive. When Write primitive is invoked, the warp directly writes the elements into the bucket according to their priority values. Differing from ADDS \cite{wang2021fast} with Single-Reader-Multiple-Writers (SRMW) design, ours adopts a Multiple-Readers-Multiple-Writers (MRMW) design to better tailor to MLMQ.

\textbf{L2 Priority queue.} The state-of-the-art heap-based GPU priority queue is BGPQ \cite{chen2021bgpq}. However, it is implemented in block-level parallelism and cannot be directly applied in MLMQ. We have provided the warp-level priority queue following the basic idea. Specifically, the priority queue consists of nodes organized as a complete binary tree. Each node contains a batch of sorted elements and a mutex lock. Before performing any operation on a node, the warp first acquires its mutex lock. The lock is released after completion. Read/Write primitives are performed with implemented operations, including node insertion, deletion, and percolation. Such a fine-grained lock-based algorithm enables parallel node operation, which leads to a theoretical increase in parallelism.

\textbf{L2 Multi-queue.} We implement the queue \cite{rihani2015multiqueues, postnikova2022multi} with L2 priority queues. Specifically, we maintain $pnum$ priority queues. Each queue is associated with a set of warps and serves as their reading queue. Warps only read elements from the reading queue, but write elements to all the $pnum$ queues. For load balancing, each warp records the queue ID of its last write operation. The next write operation is performed on the next queue in sequence. The ID starts from the ID of the reading queue and cycles back and forth.

\section{Evaluation}

\subsection{Experimental Setup}
\noindent \textbf{Platform}. We conduct our experiments on three NVIDIA GPUs: RTX 3080Ti, Tesla A100, and RTX 4090. The host program and the CPU baselines are executed on Intel Core i9-12900KF. Detailed specifications are listed in Table \ref{tab:hardware}.

\begin{table}[h]
\centering
\scalebox{0.7}{
\begin{tabular}{c|c|c|c|c|c}
    \toprule
    Hardware & Architecture & Memory & Bandwidth & FP32 Peak & Cores\\
    \midrule
    RTX 3080Ti & \multirow{2}{*}{Ampere} & GDDR6X & 912 GB/s & 34.1 TFLOPS & 80 (SMs)\\
    Tesla A100 & & HBM2e & 1,935 GB/s & 19.5 TFLOPS & 108 (SMs)\\
    \midrule
    RTX 4090 & Ada Lovelace & GDDR6X & 1,010 GB/s & 82.58 TFLOPS & 128(SMs) \\
    \midrule
    i9-12900KF & Alder Lake & DDR4 & 76.8 GB/s & 159.5 GFLOPS & 16\\ 
    \bottomrule
\end{tabular}
}
\caption{Specifications of hardware platforms.}
\label{tab:hardware}
\end{table}

\noindent \textbf{Baseline}. We compare the performance of our MLMQ-based SSSP with the following implementations: \emph{ADDS}\cite{wang2021fast}, the state-of-the-art GPU implementation of $\Delta$-stepping algorithm. \emph{H-BF}\cite{busato2015hbf}, an efficient implementation of the Bellman-Ford Algorithm on GPUs. \emph{Gunrock}\cite{wang2016gunrock}, a high-performance graph processing library on GPUs. We also use two CPU baselines in Gunrock, referred to as \emph{Gunrock-CPU}, and in Boost library, referred to as \emph{Boost-CPU} \cite{schaling2011boost}.

\begin{table}[h]
\centering
\scalebox{0.75}{
\begin{tabular}{c|c|r|r|r|r}
    \toprule
    Type & Graph name & Vertices & Edges & Avg & Std \\
    \midrule
    \multirow{8}{*}{R} & USA-road-d.NY & 264K & 730K & 2.76 & 0.98 \\
    & USA-road-d.BAY & 321K & 800K & 2.49 & 0.99 \\
    & USA-road-d.COL & 436K & 1.1M & 2.43 & 0.94 \\
    & USA-road-d.FLA & 1.1M & 2.7M & 2.53 & 0.96 \\
     & USA-road-d.CAL & 1.9M & 4.6M & 2.46 & 0.94 \\
     & USA-road-d.E & 3.6M & 8.8M & 2.44 & 0.95 \\
     & USA-road-d.W & 6.3M & 15.2M & 2.43 & 0.93 \\
     & USA-road-d.USA & 23.9M & 57.7M & 2.43 & 0.85 \\
    \midrule
    \multirow{8}{*}{MC} & delaunay\_n18 & 262K & 1.6M & 6.00 & 1.34 \\
    & delaunay\_n20 & 1.0M & 6.3M & 6.00 & 1.34 \\
    & delaunay\_n23 & 8.4M & 50M & 6.00 & 1.34 \\
    & delaunay\_n24 & 16.8M & 100.7M & 6.51 & 1.36 \\
    & G3\_circuit & 1.6M & 9.2M & 5.83 & 0.64 \\
    & memchip & 2.7M & 14.8M & 5.47 & 2.04 \\
    & Freescale1 & 3.4M & 18.9M & 5.52 & 2.06 \\
    & circuit5M\_dc & 3.5M & 19M & 5.45 & 2.09 \\
    
    \midrule
    \multirow{4}{*}{PL} & flicker &  821K & 9.8M & 11.98 & 87.16 \\
    & rmat20 & 1.0M & 8.3M & 7.88 & 16.48 \\
    & rmat22 & 4.2M & 33M & 7.92 & 18.45\\
    & Synthetic* & 1.2M $\sim$ 1.8M & 8M $\sim$ 64M & 6.6 $\sim$ 35 & 15 $\sim$ 99\\
    
    \bottomrule
\end{tabular}
}
\caption{Detailed information of graph datasets. Avg and Var indicate the average and standard deviation of vertex degrees. Graph type abbreviation: \textbf{R} for Road Networks, \textbf{MC} for graphs in Meshes and Circuit problems, and \textbf{PL} for Power-Law graphs. \textbf{*} indicates a collection of generated graphs.}
\label{tab:graph}
\end{table}

\noindent \textbf{Dataset}. We use graph datasets from different fields of applications, including road networks \cite{demetrescu2009usaroad}, 3D meshes \cite{murphy2010graph500}, circuit simulation problems \cite{davis2011university}, and social networks \cite{davis2011university}. We also use synthetic graphs, including RMAT graphs \cite{murphy2010graph500} and 8 graphs created by a power-law graph generator \cite{khorasani2015parmat}, with 1.2M $\sim$ 1.8M vertices and 8M $\sim$ 64M edges. Details of the datasets are shown in Table \ref{tab:graph}.
\par
In addition, we include 680 matrices with $>10^3$ vertices from the SuiteSparse Matrix Collection \cite{davis2011university}, to evaluate performance generalizability and to build the input-adaptive model. Those matrices can be formulated as graphs without negative edge weights and are runnable by all baselines.

\subsection{Overall Improvement}

\begin{figure}[t]
    \centering
    \includegraphics[width=\linewidth]{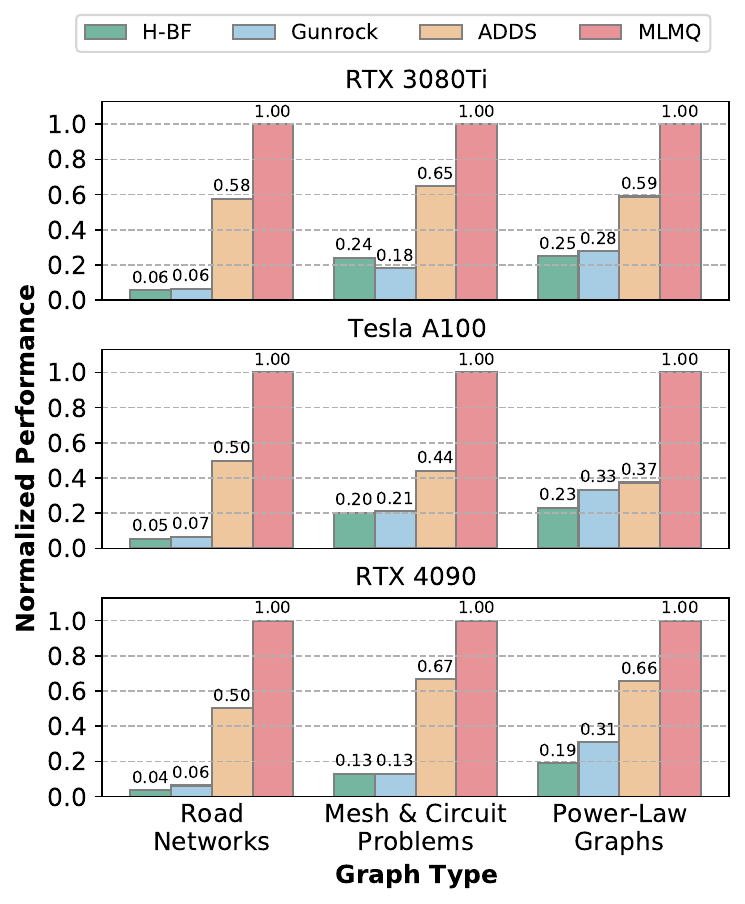}
    \caption{Performance comparison across different SSSP implementations on three GPUs. The values report the average performance over all graphs within each type.}
    \label{pic:overall}
\end{figure}

This section compares the performance across MLMQ and baselines. The results of execution time on three NVIDIA GPUs are shown in Figure \ref{pic:overall}. On RTX 3080Ti, MLMQ achieves average speedups of 1.87x over ADDS, 15.08x over H-BF, and 17.13x over Gunrock. On Tesla A100, MLMQ achieves average speedups of 2.67x over ADDS, 12.54x over H-BF and 11.31x over Gunrock. On RTX 4090, MLMQ achieves average speedups of 1.94x over ADDS, 14.92x over H-BF, and 16.81x over Gunrock. It is also worth noting that MLMQ outperforms other baselines on every individual graph. 

We compare the time of MLMQ with CPU baselines. On RTX 3080Ti, MLMQ achieves average speedups of 100.7x over Boost-CPU and 104.8x over Gunrock-CPU. On Tesla A100, the speedups are 114.5x and 125.2x, respectively. On RTX 4090, the speedups are 187.3x and 198.9x, respectively.

\begin{figure}[t]
    \centering
    \includegraphics[width=\linewidth]{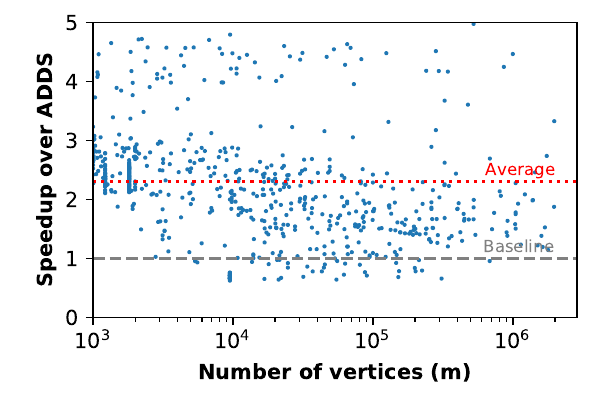}
    \caption{Speedup over SOTA baseline ADDS with 680 SuiteSparse matrices on NVIDIA RTX 3080Ti.}
    \label{pic:suite_overall}
\end{figure}

We further compare MLMQ against the strongest baseline, ADDS, on the SuiteSparse matrices. The results are shown in Figure~\ref{pic:suite_overall}. MLMQ achieves speedups on more than 91.5\% of the matrices, with an average speedup exceeding 2.3x, demonstrating its generality as the efficient SSSP solution.

\subsection{Performance Analysis}

\begin{figure}[t]
    \centering
    \includegraphics[width=\linewidth]{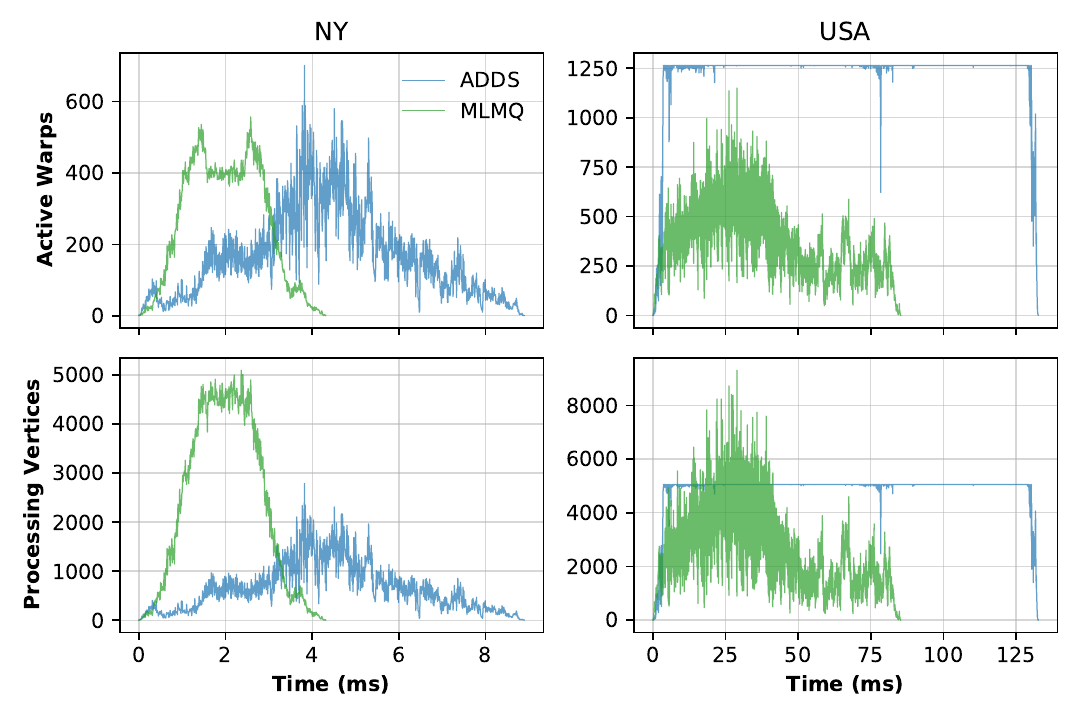}
    \caption{Runtime behavior of ADDS and MLMQ on two road networks. The first row indicates overall parallelism, and the second row indicates the workload over time.}
    \label{pic:warp_vertex}
\end{figure}

To better understand the performance improvements, we sampled and compared ADDS (the state-of-the-art SSSP implementation) and MLMQ during kernel execution, measuring the number of active warps (degree of parallelism) and processing vertices (work efficiency) over time. The sampling results are shown in Fig.~\ref{pic:warp_vertex}.

On the small road network (NY), MLMQ significantly improves parallelism by 1.53x, at a cost of 1.7x redundant work. Despite the larger workload, the higher parallelism and more efficient queue access can offset the extra overhead, resulting in a 2.06x speedup. In contrast, on the large road network (USA), MLMQ trades parallelism for better work efficiency. MLMQ reduces the average active warps to 0.27x of ADDS, but cuts the total workload by 3.5x, resulting in a 1.56x speedup. The analysis shows MLMQ's better ability to achieve an ideal balance between parallelism and work efficiency.

\begin{figure*}[t]
    \centering
    \includegraphics[width=\linewidth]{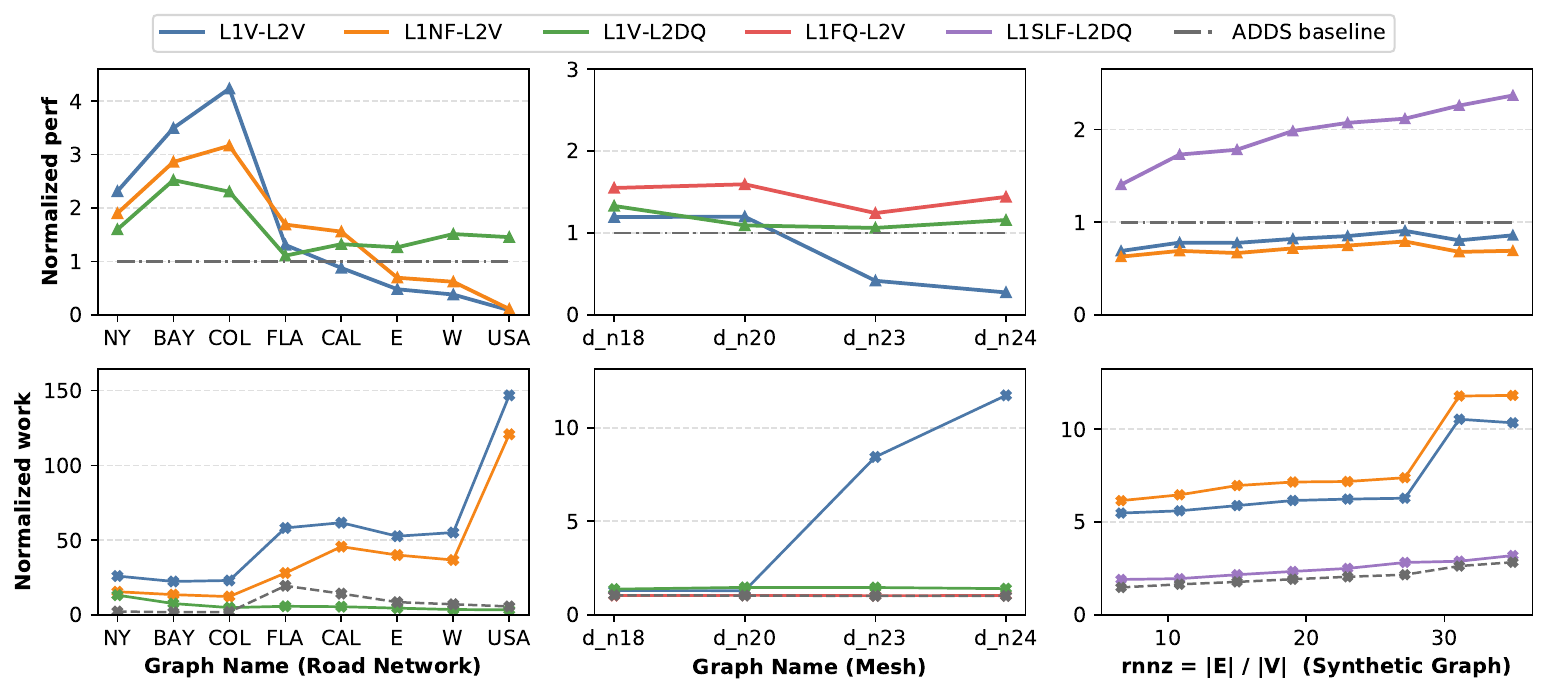}
    \caption{Comparison of normalized performance and total work across different graph types.}
    \label{pic:all}
\end{figure*}

\subsection{MLMQ Configuration Analysis}
\label{sec:test case}
We compare the execution time and work efficiency of different MLMQ configurations across three categories of graphs: road networks, meshes/circuits, and power-law graphs. Figure~\ref{pic:all} summarizes the results, where performance is normalized to ADDS~\cite{wang2021fast}, and total work is normalized by the number of vertices. Configurations are denoted as L1*-L2*, where L1* and L2* represent the choices of L1 and L2 queues, respectively. L0 queues are always vectors.

Road networks feature small degrees and large diameters, thus exhibiting high theoretical parallelism. Configurations with higher parallelism (L1V-L2V on small graphs, L1NF-L2V on medium graphs) achieve better performance, while larger graphs require the more work-efficient L1V-L2DQ.  
Meshes and circuit graphs have larger degrees and smaller diameters. Here, L1FQ-L2V achieves the best balance: the L1 filter reduces priority inversion, and the L2 FIFO ensures low overhead, outperforming both L1V-L2V and L1V-L2DQ.  
Power-law graphs have skewed degree distributions and small diameters. Work efficiency dominates performance: L1SLF-L2DQ, with both priority-aware queues, reduces redundant work and warp divergence, achieving nearly 2x speedup over ADDS, especially on denser graphs.

In summary, the choice of MLMQ configuration is closely tied to both graph structure and scale, highlighting the necessity of adapting queue designs to workload characteristics.



\subsection{Input Adaptability}

\begin{figure}[t]
\centering
\includegraphics[width=0.9\linewidth]{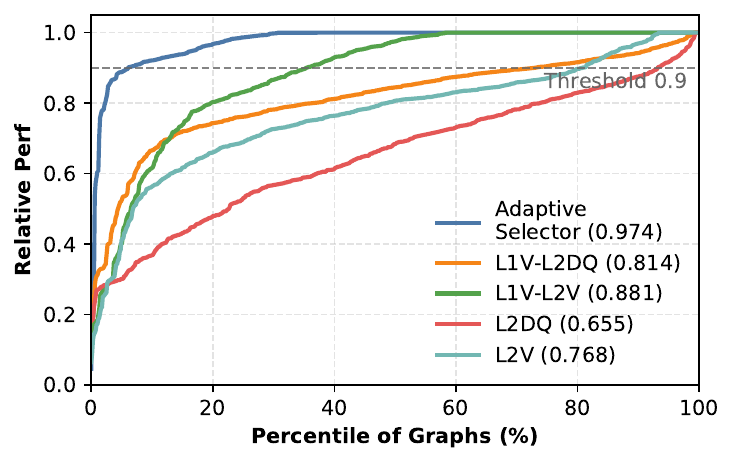}
\caption{CDF of relative performance measured on SuiteSparse matrices. Legends indicate the average performance values (shown in parentheses).}
\label{pic:sel_quality_val}
\end{figure}

In this section, we evaluate the relative performance of different queue selections on the SuiteSparse benchmark. Relative performance is defined as the ratio between the runtime of a given method on a graph and the best runtime among all queue types and parameter configurations on the same graph. The CDF curve shows how many graphs can achieve at least a given level of relative performance. The selector is trained on 70\% of the graphs and validated on the remaining 30\%. The results in Figure~\ref{pic:sel_quality_val} show two key findings. First, MLMQ-based designs (L1V-L2V, L1V-L2DQ) consistently outperform traditional single-queue baselines, yielding average improvements of 1.15x–1.35x. Second, our adaptive selector further improves over the fixed configuration by over 1.1$\times$ on average. Moreover, to measure how many graphs can reach near-optimal performance, we set a threshold of 0.9. Under this criterion, our method achieves 93.9\% coverage, while the best fixed queue (L1V-L2V) reaches only 64.6\%. These results demonstrate both the necessity and the effectiveness of the proposed input-adaptive configuration method.

\subsection{Case Study 1: Priority Queues}

\begin{figure}[t]
    \centering
    \includegraphics[width=0.95\linewidth]{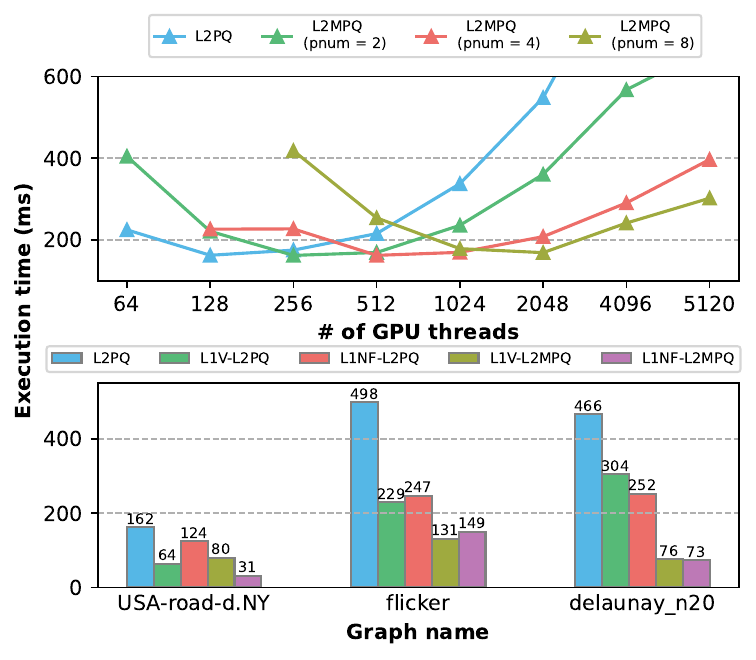}
    \caption{Upper: Performance of L2PQ and L2MPQ at different concurrency levels on the graph \codeemph{USA-road-d.NY}. Lower: Performance trends when using different L1 queues.}
    \label{pic:priorityq}
\end{figure}

We discuss the performance of priority queues (L2PQ, \cite{chen2021bgpq}) and multi-queues (L2MPQ, \cite{rihani2015multiqueues, postnikova2022multi}). Despite being widely used on CPUs, those queues exhibit poor performance on GPUs. 
The upper part of Figure \ref{pic:priorityq} shows that the performance of L2PQ and L2MPQ varies with the number of threads. Using L2MPQ alleviates the contention under high concurrency. However, it does not lead to improvement under the optimal parallelism configuration, since the bottleneck at the root nodes remains unresolved. In the lower part of Figure \ref{pic:priorityq}, we introduce L0 and L1 queues to alleviate the bottleneck, which leads to speedups of up to 6.40x. Based on the above results, we point out that the CPU-effective multi-queue solutions cannot be directly applied to GPUs. MLMQ leads to a better solution. However, with the high global access costs present, priority queues are not competitive in MLMQ compared to the GPU-friendly L2 queues like L2DQ and L2V.


\subsection{Case Study 2: BFS Results}

\begin{figure}[t]
    \centering
    \includegraphics[width=0.95\linewidth]{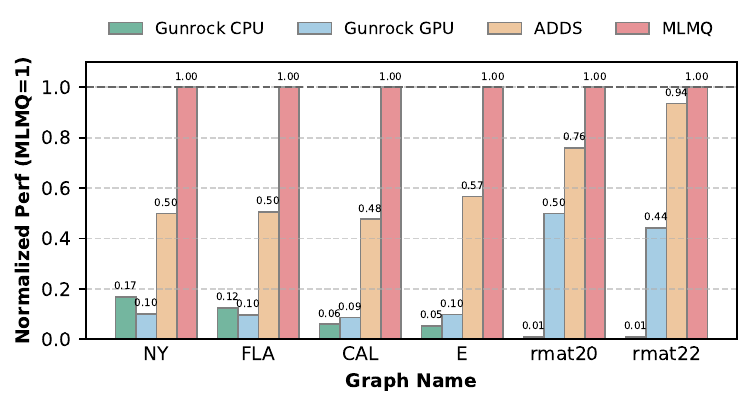}
    \caption{Performance comparison across different graph frameworks on BFS workload.}
    \label{pic:bfs}
\end{figure}

To further demonstrate the generality of the MLMQ structure, we evaluate its performance on the Breadth-First Search (BFS) problem, the most fundamental graph traversal algorithm~\cite{cormen2022introduction}. Due to its wide application in navigation and relationship analysis, we use road networks and power-law graphs as representative inputs and compare MLMQ against state-of-the-art GPU and CPU frameworks. As shown in Figure~\ref{pic:bfs}, MLMQ consistently outperforms the baselines. On average, MLMQ achieves 43.26x, 7.74x, and 1.71x speedups over Gunrock-CPU, Gunrock-GPU, and ADDS, respectively. These results confirm that MLMQ can effectively generalize to other graph algorithms while maintaining strong performance. This generality stems from its algorithm-independent design and the modularity of its queue framework.

\section{Related Work}
\textbf{SSSP on GPUs}. As a classic graph-based problem, SSSP faces great challenges in implementing on GPUs due to its irregular pattern and serial nature. Existing implementations are based on Dijkstra's algorithm \cite{martin2009relatedcuda, ortega2013relatednew}, Bellman-ford algorithm \cite{burtscher2012relatedquantitative, busato2015hbf, surve2017relatedparallel} and $\Delta$-stepping algorithm \cite{davidson2014work, wang2021fast}. However, a single and fixed queue design cannot consistently achieve good performance on various input graphs. MLMQ provides a broader optimization space by constructing novel data structures while also encompassing existing queue designs.

\noindent \textbf{Concurrent queues}. In graph algorithms, queues are used for task scheduling. The parallelism and work efficiency of the queue are critical. Existing concurrent queue designs include priority queues \cite{chen2021bgpq, crosetto2019cupq, he2012design, alistarh2015spraylist, moscovici2017gpu}, MultiQueues \cite{rihani2015multiqueues, postnikova2022multi}, GWLR queues \cite{jeffrey2015scalable}, Global-local queues \cite{yesil2019understanding, nguyen2013lightweight, nguyen2011synthesizing}, etc. Those designs aim to increase parallelism, reduce contention, or enhance memory access efficiency. 
\par
\noindent \textbf{Graph processing frameworks}. Graph processing frameworks provide efficient interfaces for typical graph operations, in order to reduce the manual costs of graph algorithm implementation. Frameworks on GPUs include GTS \cite{kim2016gts}, Groute \cite{ben2017groute}, Gunrock \cite{wang2016gunrock}, Tigr \cite{nodehi2018tigr}, etc. 
These frameworks focus more on the ease of programming while falling short of the optimal performance.

\section{Conclusion}

SSSP is one of the most important problems in graph algorithms. To efficiently solve SSSP on GPUs, we propose MLMQ, a novel data structure design that addresses the concurrency contention and global memory access overheads inherent in traditional single-queue design. We implement MLMQ with a cache-like queue collaboration mechanism that enables a modular queue framework, upon which we design GPU-friendly and input-adaptive solutions for SSSP. Experiments on real-world graphs demonstrate the effectiveness (over \lowspeedup x improvement) and robustness of MLMQ.


Currently, the scope of MLMQ is limited to the SSSP and BFS-style traversal algorithms. In future work, we plan to extend MLMQ to more graph algorithms. As MLMQ follows a unified and algorithm-independent design, such extensions are achievable with appropriate engineering efforts. Additionally, one of our future directions is to extend MLMQ to distributed scenarios by introducing a multi-GPU–shared L3 queue maintained through interconnections.



\bibliographystyle{ACM-Reference-Format}   
\bibliography{ref}

\end{document}